PREPRINT ARTICLE

# The server is dead, long live the server: Rise of Serverless Computing, Overview of Current State and Future Trends in Research and Industry

Paul Castro[1], Vatche Ishakian[2], Vinod Muthusamy[1], Aleksander Slominski[1]

[1] IBM T.J. Watson Research Center, {castrop, vmuthus, aslom}@us.ibm.com
[2] Computer Information Systems, Bentley University, vishakian@bentley.edu

## Abstract

Serverless computing -- an emerging cloud-native paradigm for the deployment of applications and services -- represents an evolution in cloud application development, programming models, abstractions, and platforms. It promises a real pay-as-you-go billing (with millisecond granularity) with no waste of resources, and lowers the bar for developers by asking them to delegate all their operational complexity and scalability to the cloud provider. Delivering on these promises comes at the expense of restricting functionality. In this article we provide an overview of serverless computing, its evolution, general architecture, key characteristics and uses cases that made it an attractive option for application development. Based on discussions with academics and industry experts during a series of organized serverless computing workshops (WoSC),  we also identify the technical challenges  and open problems.

## Introduction

Cloud computing in general, and Infrastructure-as-a-Service (IaaS) in particular, have become widely accepted and adopted paradigms for computing with the offerings of Virtual Machines (VM) on demand. By 2020, 67% of enterprise IT infrastructure and software spending will be for cloud-based offerings [1].

A major factor in the increased adoption of the cloud by enterprise IT was its pay-as-you-go model where a customer pays only for resources leased from the cloud provider and have the ability to get as many resources as needed with no up-front cost (elasticity) [20]. Unfortunately, the burden of scaling was left for developers and system designers that typically used over-provisioning techniques to handle sudden surges in service requests. Studies of reported usage of cloud resources in data centers [21], show a substantial gap between the resources that cloud customers allocate and pay for (leasing VMs), and actual resource utilization (CPU, memory, etc.).

Serverless computing is emerging as a new and compelling paradigm for the deployment of cloud applications, largely due to the recent shift of enterprise application architectures to containers and microservices [2]. Using serverless gives pay-as-you-go without additional work to start and stop server and is closer to original expectations for cloud computing to be treated like as a utility [20]. Developers using serverless computing can get cost saving and scalability without need to have high level of cloud computing expertise that is time-consuming to acquire.

Due to its simplicity and economical advantages serverless computing is gaining popularity as reported by the increasing rate of the "serverless" search term by Google Trends. Its market size is estimated to grow to 7.72 billion by 2021 [3]. Most prominent cloud providers including Amazon, IBM, Microsoft, and Google have already released serverless computing capabilities with several additional open-source efforts driven by both industry and academic institutions (for example see CNCF Serverless Cloud Native Landscape[1].

From the perspective of an Infrastructure-as-a-Service (IaaS) customer, the serverless paradigm shift presents both an opportunity and a risk. On the one hand, it provides developers with a simplified programming model for creating cloud applications that abstracts away most, if not all, operational concerns. They no longer have to worry about availability, scalability, fault tolerance, over/under provisioning of VM resources, managing servers and other infrastructure issues, and instead focus on the business aspects of their applications. The paradigm also lowers the cost of deploying cloud code by charging for execution time rather than resource allocation. On the other hand, deploying such applications in a serverless platform is challenging and requires relinquishing design decisions to the platform provider that concern, among other things, quality-of-service (QoS) monitoring, scaling, and fault-tolerance properties. There is a risk an application's requirements may evolve to conflict with the capabilities of the platform.

---

[1] https://s.cncf.io/

From the perspective of a cloud provider, serverless computing is an additional opportunity to control the entire development stack, reduce operational costs by efficient optimization and management of cloud resources, offer a platform that encourages the use of additional services in their ecosystem, and lower the effort required to author and manage cloud-scale applications.

## Defining serverless computing

Serverless computing can be defined by its name - less thinking (or caring) about servers. Developers do not need to worry about low-level details of servers management and scaling, and only pay for when processing requests or events. We define serverless as follows.

**Definition: serverless computing is a cloud-native platform that hides server usage from developers and runs developer code on-demand automatically scaled and billed only for the time the code is running.**

This definition captures the two key features of serverless computing:

1. **Cost - billed only for what is running (pay-as-you-go)**
   As servers and their usage is not part of serverless computing model then it is natural to pay only for when code is running and not for idle servers. As execution time may be short the it should be charged in fine grained time units (like 100s of milliseconds) and developers do need to pay for overhead of servers creation or destructions (such as VM booting time). This cost model is very attractive to workloads that need to run occasionally - serverless essentially supports "scaling to zero" and avoid need to pay for idle servers. The big challenge for cloud providers is the need schedule and optimize cloud resources.
2. **Elasticity - scaling from zero to "infinity"**
   Since developers do not have control over servers that run their code, nor do they know the number of servers their code runs on, decisions about scaling are left to cloud providers. Developers do not need to write auto-scaling policies or define how machine level usage (CPU, memory. etc.) translates to application usage. Instead they depend on the cloud provider to automatically start more parallel executions when there is more demand for it. Developers also can assume that cloud provider will take care of maintenance, security updates, availability and reliability monitoring of servers.

Serverless computing today typically favors small, self contained units of computation to make it easier to manage and scale in the cloud. A computation can not depend on the cloud platform to maintain state which can be interrupted or restarted, which inherently influences the serverless computing programming models. There is, however, no equivalent notion of scaling to zero when it comes to state, since a persistent storage layer is needed. However, even if the implementation of a stateful service requires persistent storage, a provider can offer a pay-as-you-go pricing model that would make state management serverless.

The most natural way to use serverless computing is to provide a piece of code (function) to be executed by the serverless computing platform. It leads to the rise of Function-as-a-service (FaaS) platforms focused on allowing small pieces of code represented as functions to run for limited amount of time (at most minutes), with executions triggered by events or HTTP requests (or other triggers), and not allowed to keep persistent state (as function may be restarted at any time). By limiting time of execution and not allowing functions to keep persistent state FaaS platforms can be easily maintained and scaled by service providers. Cloud providers can allocate servers to run code as needed and can stop servers after functions finish as they run for limited amount of time. If functions need to maintain state then they can use external services to persist their state.

FaaS is an embodiment of serverless computing principles, which we define as follows.

**Definition: Function-as-a-Service (FaaS) is a serverless computing platform where the unit of computation is a function that is executed in response to triggers such as events or HTTP requests.**

Our approach to defining serverless is consistent with emerging definitions of serverless from industry. For example, Cloud Native Computing Foundation (CNCF) defines serverless computing [25] as *"the concept of building and running applications that do not require server management. It describes a finer-grained deployment model where applications, bundled as one or more functions, are uploaded to a platform and then executed, scaled, and billed in response to the exact demand needed at the moment."* . While our definition is close to the CNCF definition, we make a distinction between serverless computing and providing functions as unit of computation. As we discuss in the research challenges section, it is possible that serverless computing will expand to include additional aspects that go beyond today's relatively restrictive stateless functions into possibly long-running and stateful execution of larger compute units. However, today serverless and FaaS are often used interchangeably as they are close in meaning and FaaS is the most popular type of serverless computing.

Paul Johnston (ServerlessDays co-founder) defined serverless as follows[2]: "A Serverless solution is one that costs you nothing to run if nobody is using it (excluding data storage)". This definition highlights the most important characteristic of serverless computing - pays-as-you-go. It assumes serverless computing is subset of cloud computing so auto-scaling is included and as servers are not mentioned means that developers have no access to servers. CNCF and our definitions emphasize not only pay-as-you-go or "scale to zero" aspects but also emphasize the no need to manage servers.

Another way to define serverless computing is by what functionality it enables. Such approach emphasizes "serverless is really about the managed services" and FaaS can be treated as cloud "glue" as described by Steven Faulkner (a senior software engineer at LinkedIn)[3] - it is "glue" as it keeps together application composed out of cloud services. Such definition addresses only a narrow set of use cases where serverless computing is used, while our definition captures the important uses cases which we will highlight in the use cases section.

All definitions share the observation that the name 'serverless computing' does not mean servers are not used, but merely that developers can leave most operational concerns of managing servers and other resources, including provisioning, monitoring, maintenance, scalability, and fault-tolerance to the cloud provider.

---

[2] https://medium.com/@PaulDJohnston/a-simple-definition-of-serverless-8492adfb175a
[3] https://read.acloud.guru/serverless-is-eating-the-stack-and-people-are-freaking-out-and-they-should-be-431a9e0db482

## History and Related Work

The term 'serverless' can be traced to its original meaning of not using servers and typically referred to peer-to-peer (P2P) software or client side only solutions [15]. In the cloud context, the current serverless landscape was introduced during an AWS re:Invent event in 2014 [10]. Since then, multiple cloud providers, industrial, and academic institutions have introduced their own serverless platforms. Serverless seems to be the natural progression following recent advancements and adoption of VM and container technologies, where each step up the abstraction layers led to more lightweight units of computation in terms of resource consumption, cost, and speed of development and deployment. Furthemore, serverless builds upon long-running trends and advances in both distributed systems, publish-subscribe systems, and event-driven programming models [11], including actor models [12], reactive programming [13], and active database systems [14].

Serverless platforms can be considered as evolution of Platform-as-a-Service (PaaS) as provided by platforms such as Cloud Foundry, Heroku, and Google App Engine (GAE). The main differentiators of serverless platforms, as compared to previous PaaS, include transparent autoscaling and fine-grained resource charging only when code is running. That should not to be confused with free usage quota where limited monthly resource quota is available but counted even if the application is not used. For example GAE Standard is priced in "instance hours"[4] and even if the app is not used the instance is kept running. Later GAE added Flexible version with more fine grained billing unit but still developers will be billed by even if server is not used. That can lead to unexpected outcomes when the bill arrives at the end of month for forgotten test services[5].

Mobile Backend as-a-Service (MBaaS) or more generalized Backend as-a-Service (BaaS) bears a close resemblance to serverless computing. Some of those services even provided "cloud functions" (e.g., Facebook's now-defunct Parse Cloud Code). Such code, however, was typically limited to mobile use cases.

---

[4] https://cloud.google.com/appengine/docs/the-appengine-environments
[5] https://stackoverflow.com/questions/47125661/

Software-as-a-Service (SaaS) may support the server-side execution of user provided functions but they are executing in the context of an application and hence limited to the application domain. Some SaaS vendors allow the integration of arbitrary code hosted somewhere else and invoked via an API call. For example, this is approach is used by the Google Apps Marketplace in Google Apps for Work.

|  | **IaaS** | **PaaS** | **Serverless (FaaS)** | **BaaS/SaaS** |
| --- | --- | --- | --- | --- |
| **Expertise required** | High | Medium | Low | Low |
| **Developer Control / Customization allowed** | High | Medium | Low | Very low |
| **Scaling / Cost** | Requires high-level of expertise to build auto-scaling rules and tune them | Requires high-level of expertise to build auto-scaling rules and tune them | Auto-scaling to work load requested (function calls), and only paying for when running (scale to zero) | Hidden from users, limits set based on pricing and QoS |
| **Unit of work deployed** | Low-level infrastructure building blocks (VMs, network, storage) | Packaged code that is deployed and running as a service | One function execution | App-specific extensions |
| **Granularity of billing** | Medium to large granularity: minutes to hours per resource to years for discount pricing | Medium to large granularity: minutes to hours per resource to years for discount pricing | Very low granularity: 100s of milliseconds of function execution time | Large: typically subscription available based on maximum number of users and billed in months |

**Table 1**: Comparison of different choices for cloud as a service

The boundaries defining serverless computing functionality overlaps with Platform-as-a-Service (PaaS) and Software-as-a-Service (SaaS). One way to categorize serverless is to consider the varying levels of developer control over the infrastructure. In an Infrastructure-as-a-Service (IaaS) model, the developer has much more control over the resources, but is responsible for managing both the application code and operating the infrastructure. This gives the developer great flexibility and the ability to customize every aspect of the application and infrastructure, such as administering VMs, managing capacity and utilization, sizing the workloads, achieving fault tolerance and high availability. PaaS abstracts away VMs and takes care of managing underlying operating systems and capacity, but the developer is responsible for the full lifecycle of the code that is deployed and run by the platform, which does not scale down to zero. SaaS represent the other end of the spectrum where the developer has no control over the infrastructure, and instead get access to prepackaged components. The developer is allowed to host code there, though that code may be tightly coupled to the platform. Backend-as-a-Service (BaaS) is similiar to SaaS in that the functionality is targeting specific use cases and components, for example Mobile Backend-as-a-Service (MBaaS) provide backend functionality needed for mobile development such as managing push notifications, and when it allows developer to run code it is within that backend functionality (see Table 1).

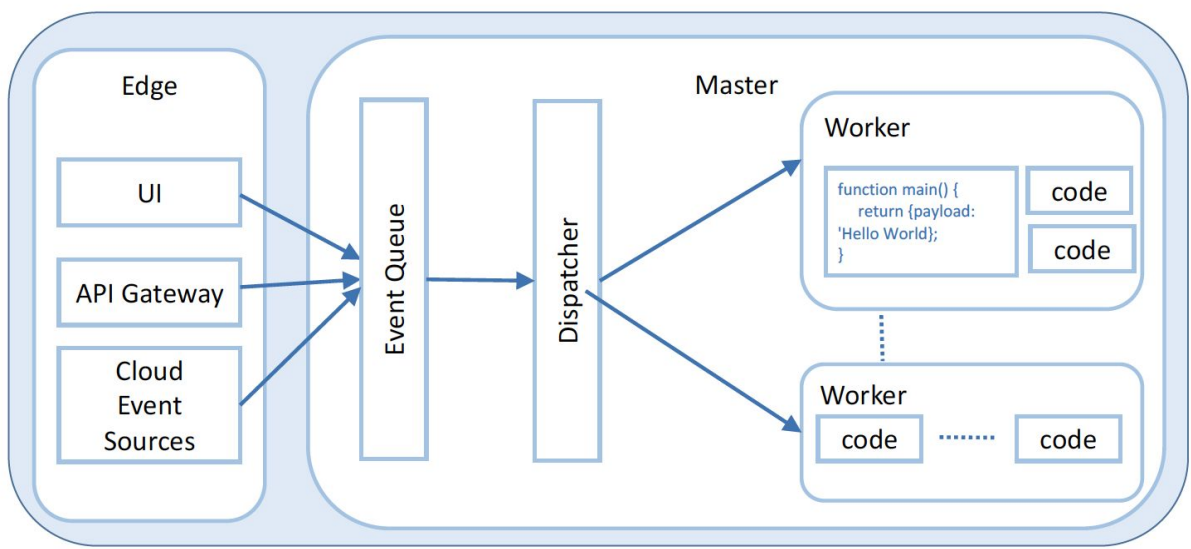

Figure 1: High level serverless platform architecture

## Architecture

The core functionality of a serverless framework is simply that of an event processing system, as shown in Figure 1. The service manages a set of used defined functions (a.k.a actions). Once a request is received over HTTP from an event data source (a.k.a. triggers), the system determines which action(s) should handle the event, create a new container instance, send the event to the function instance, wait for a response, gather execution logs, make the response available to the user, and stop the function when it is no longer needed.

The abstraction level provided by FaaS is unique: a short running stateless function. This has proven to be both expressive enough to build useful applications but simple enough to allow the platform to autoscale in an application agnostic manner.

While the architecture is relatively simple, the challenge is to implement such functionality while considering metrics such as cost, scalability, latency, and fault tolerance. To isolate the execution of functions from different users in a multi-tenant environment, container technologies [22], such as Docker, are often used.

Upon the arrival of an event, the platform proceeds to validate the event making sure that it has the appropriate authentication and authorization to execute. It also checks the resource limits for that particular event. Once the event passes validation, the platform the event is queued to be processed. A worker fetches the request, allocates the appropriate container, copies over the function -- use code from storage -- into the container and executes the event. The platform also manages stopping and deallocating resources for idle function instances.

Creating, instantiating, and destroying a new container for each function invocation while can be expensive, and introduces an overall latency which is referred to as the cold start problem. In contrast, warm containers are containers that were already instantiated and executed a function. Cold start problems can be mitigated by techniques such as maintaining a pool of uninstantiated stem cell containers, which are containers that have been previously instantiated but not assigned to a particular user, or reuse a warm container that have been previously invoked for the same user [23]. Another factor that can affect the latency is the reliance of the user function on particular libraries (e.g. numpy) that need to be downloaded and installed before function invocation. To reduce startup time of cloud functions, one can appropriately cache the most important packages across the node workers thus leading to reduced startup times [29].

In typical serverless cloud offerings, the only resource configuration customers are allowed to configure is the size of main memory allocated to a function. The system will allocate other computational resources (e.g., CPU) in proportion to the main memory size. The larger the size the higher the cpu allocation. Resource usage is measured and billed in small increments (e.g.,100ms) and users pay only for the time and resources used when their functions are running.

Several open source serverless computing frameworks are available from both industry and academia (e.g. Kubeless, OpenLambda, OpenWhisk, OpenFaaS). In addition, major cloud vendors such as Amazon, IBM, Google, and Microsoft have publically available commercial serverless computing frameworks for their consumers. While the general properties (e.g. memory, concurrent invocations, maximum execution duration of a request) of these platforms are relatively the same, the limits as set by each cloud provider are different. Note the limits on these properties are a moving target and are constantly changing as new features and optimizations are adopted by cloud providers. Evaluating the performance of different serverless platform to identify the tradeoffs has been a recent topic of investigation [17, 26, 27], and recent benchmarks have been developed to compare the serverless offering by the different cloud providers[6].

## Programming Model

A typical serverless programming model consists of two major primitives: Action and *Trigger*. An *Action* is a stateless function that executes arbitrary code. Actions can be invoked asynchronously in which the invoker -- caller request -- does not expect a response, or synchronously where the invoker expects a response as a result of the action execution. A *Trigger* is a class of events from a variety of sources. Actions can be invoked directly via REST API, or executed based on a trigger. An event can also trigger multiple functions (parallel invocations), or the result of an action could also be a trigger of another function (sequential invocations). Some serverless frameworks provide higher level programming abstractions for developers such as function packaging, sequencing, and composition which may make it easier to construct more complex serverless apps.

---

[6] http://faasmark.com/

Currently, serverless frameworks execute a single main function that takes a dictionary (such as a JSON object) as input and produces a dictionary as output. They have limited expressiveness as they are built to scale. To maximize scaling, serverless functions do not maintain state between executions. Instead, the developer can write code in the function to retrieve and update any needed state. The function is also able to access a context object that represents the environment in which the function is running (such as a security context). As shown in the example below, a function written in JavaScript could take as input a JSON object as the first parameter, and context as the second.

```
function main(params, context) {
   return {payload:  'Hello, ' + params.name
                   + ' from ' + params.place};
}
```

Current Cloud Provider Serverless offerings support a wide variety programming languages, including Java, Python, Swift, C#, and Node.js. Some of the platforms also support extensibility mechanisms for code written in any language as long as it is packaged in a Docker image that supports a well-defined API.

Due to the limited and stateless nature of serverless functions, and its suitability for composition of APIs, cloud providers are offering an ecosystem of value added services that support the different functionalities a developer may require, and is essential for production ready applications. For example, a function may need to retrieve state from permanent storage, such as a file server or database, another may use a machine learning service to perform some text analysis or image recognition. While the functions themselves may scale due to the serverless guarantees, the underlying storage system itself must provide reliability and QoS guarantees to ensure smooth operation.

## Tools and Frameworks

One of the major challenges that is slowing the adoption of serverless is the lack of tools and frameworks. The tools and frameworks currently available can be categorized as follows: development, testing, debugging, deployment. Several solutions been proposed to deal with these categories.

Almost all cloud providers provide a cloud based IDE, or extensions/plugins to popular IDEs that allows the developer to code and deploy serverless functions. They also provide a local containerized environment with an SDK that allows the developer to develop and test locally serverless functions before deploying it in a cloud setting. To enable debugging, function execution logs are available to the developer and recent tools such as AWS X-Ray[7] allow developers to detect potential causes of the problem [28]. Finally, there are open source frameworks[8] that allow developers to define serverless functions, triggers, and services needed by the functions. Theses frameworks will handle the deployment of these functions to the cloud provider.

## Use Cases

Serverless computing has been utilized to support a wide range of applications. From an infrastructure perspective, serverless and more traditional architectures may be used interchangeably or in combination. The determination of when to use serverless will likely be influenced by other non-functional requirements such as the amount of control over operations required, cost, as well as application workload characteristics.

From a cost perspective, the benefits of a serverless architecture are most apparent for bursty [5,6,30], compute intensive [7,8] workloads. Bursty workloads fare well because the developer offloads the elasticity of the function to the platform, and just as important, the function can scale to zero, so there is no cost to the consumer when the system is idle. Compute intensive workloads are appropriate since in most platforms today, the price of a function invocation is proportional to the running time of the function. Hence, I/O bound functions are paying for compute resources that they are not fully taking advantage of. Other options to run I/O bound functions such as a multi-tenant server application that multiplexes requests may be cheaper.

| Where is serverless used? | What do they use serverless computing for? |
|---|---|

---

[7] https://aws.amazon.com/xray/
[8] https://serverless.com/

| Aegex | Xamarin application that customers can use to monitor real-time sensor data from IoT devices.[9] |
|---|---|
| Abilisense | Manages an IoT messaging platform for people with hearing difficulties. They estimated they could handle all the monthly load for less than $15 a month[10] |
| A Cloud Guru | Uses functions to perform protected actions such as payment processing and triggering group emails. IN 2017 they had around 200K users and estimated $0.14 to deliver video course to a user[11] |
| Coca-Cola | Serverless Framework is a core component of The Coca-Cola Company's initiative to reduce IT operational costs and deploy services faster. [12] |
| Expedia | Expedia did "over 2.3 billion Lambda calls per month" back in December 2016. That number jumped four and a half times year-over-year in 2017 (to 6.2 billion requests) and continues to rise in 2018.[13] |
| Glucon | Serverless Mobile backend to reduce client app code size and avoid disruptions[14] |
| Heavywater Inc | Runs website and training courses using serverless (majority of cost per user is not serverless but storage of video). Serverless reduced their costs by 70%.[15] |
| iRobot | Backend for iRobot products [16] |
| Postlight | Mercury Web Parser is a new API from Postlight Labs that extracts meaningful content from web pages. Serving 39 Million Requests for $370/Month, or: How We Reduced Our Hosting Costs by Two Orders of Magnitude [17] |
| PyWren | Map-reduce style framework for highly parallel analytics workloads[18] |

---

[9] https://microsoft.github.io/techcasestudies/azure%20app%20service/azure%20functions/iot/mobile%20application%20development%20with%20xamarin/2017/06/05/Aegex.html
[10] https://thenewstack.io/ibms-openwhisk-serverless/
[11] https://gotochgo.com/2017/sessions/61
[12] https://www.forbes.com/sites/janakirammsv/2016/10/12/why-enterprises-should-care-about-serverless-computing/
[13] https://www.theregister.co.uk/2018/05/11/lambda_means_game_over_for_serverless/
[14] https://gluonhq.com/simplifying-mobile-apps-using-serverless-approach-case-study/
[15] https://read.acloud.guru/how-going-serverless-helped-us-reduce-costs-by-70-255adb87b093
[16] https://aws.amazon.com/solutions/case-studies/irobot/
[17] https://trackchanges.postlight.com/serving-39-million-requests-for-370-month-or-how-we-reduced-our-hosting-costs-by-two-orders-of-edc30a9a88cd
[18] http://pywren.io/

| WeatherGods | A mobile weather app that uses serverless as backend[19] |
| Santander Bank | Electronic check processing. Less than 2 dollars to process all paper checks within a year. [20] |
| Financial Engines | Mathematical calculations for evaluation and optimization of investment portfolios. 94% savings on cost approximately 110K annually.[21] |

**Table 2**: Real world applications that use serverless computing

There are many areas where serverless computing is used today. Table 2 provides a representative list of different types of applications used in different application domains along with a short description. We emphasize that this is a non exhaustive list which we use to identify and discuss emerging patterns. Interested readers can find examples by going through additional use cases that are publically available by cloud providers.

From a programming model perspective, the stateless nature of serverless functions lends themselves to application structure similar to those found in functional reactive programming. This includes applications that exhibit event-driven and flow-like processing patterns (see sidebars with Use Case 1 of thumbnail creation).

As a comparison, consider an equivalent solution implemented as an application running on a set of provisioned VMs. The logic in the application to generate the thumbnails is relatively straightforward, but the user needs to manage the VMs, including monitoring traffic loads, auto-scaling the application, and managing failures. There is also a limit to how quickly VMs can be added in response to bursty workloads, forcing the user to forecast workload patterns and pay for pre-provisioned resources. The consequence is there will always be idle resources, and it is impossible to scale down to zero VMs. In addition, there needs to be a component that monitors for changes to the S3 folder, and dispatch these change events to one of the application instances. This dispatcher itself needs to be fault-tolerant and auto-scale.

---

[19] https://thenewstack.io/ibms-openwhisk-serverless/
[20] https://www.slideshare.net/OpenWhisk/ibm-bluemix-openwhisk-serverless-conference-2017-austin-usa-the-journey-continues-whats-new-in-openwhisk-land
[21] https://aws.amazon.com/solutions/case-studies/financial-engines/

Another class of applications that exemplify the use of Serverless is composition of a number of APIs, controlling the flow of data between two services, or simplify client-side code that interacts by aggregating API calls (see sidebar Use Case 2).

Serverless computing may also turn out to be useful for scientific computing. Having ability to run functions and do not worry about scaling and paying only for what is used can be very good for computational experiments. One class of applications that started gaining momentum are compute intensive applications [8]. Early results show (see Use Case 3 in sidebar) that the performance achieved is close to specialized optimized solutions and can be done in an environment that scientists prefer such as Python.

Many "born in cloud" companies build their services to take full advantage of cloud services. Whenever possible they use existing cloud services and built their functionality using serverless computing. Before serverless computing they would need to use virtual machines and create auto-scaling policies. Serverless computing with its ability to scale to zero and almost infinite on-demand scalability allows them to focus on putting business functionality in serverless functions instead of becoming experts in low-level cloud infrastructure and server management (see Use Case 4 in sidebar for more details).

## Use Case 1: Event processing

One class of applications that exemplify the use of Serverless is event-based programming. The following use case shows an example of a bursty, compute intensive workload was popularized by AWS Lambda, and has become the "Hello, World" of serverless computing, is a simple image processing event handler function.

Netflix uses serverless functions to process video files[22]. The videos are uploaded Amazon S3 [2], which emits events that trigger Lambda functions that split the video and transcode them in parallel to different formats. The flow is depicted in the Figure 2 below.

The function is completely stateless and idempotent which has the advantage that in the case of failure (such as network problems accessing the S3 folder), the function can be executed again with no side effects.

While the example above is relatively simple, by combining serverless functions with other services from the cloud provider, more complex applications can be developed e.g. stream processing, Filtering and transforming data on the fly, chatbots, and web applications.

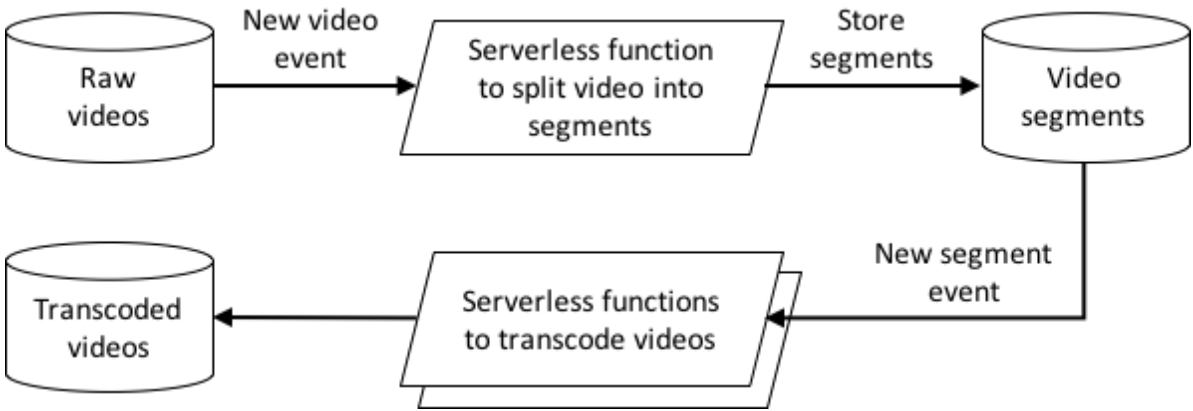

Figure 2: Video processing

---

[22] https://aws.amazon.com/solutions/case-studies/netflix-and-aws-lambda/

# Sidebar: Use case 2: API Composition

Consider a mobile app (c.f. Figure 3) that sequentially invokes a geo-location, weather, and language translation APIs to render the weather forecast for a user's current location. A short serverless function can be used to invoke these APIs. Thus the mobile app avoids invoking multiple APIs over a potentially resource constrained mobile network connection, and offloads the filtering and aggregation logic to the backend. Glucon for example, used serverless in its conference scheduler application to minimize client code, and avoid disruptions.

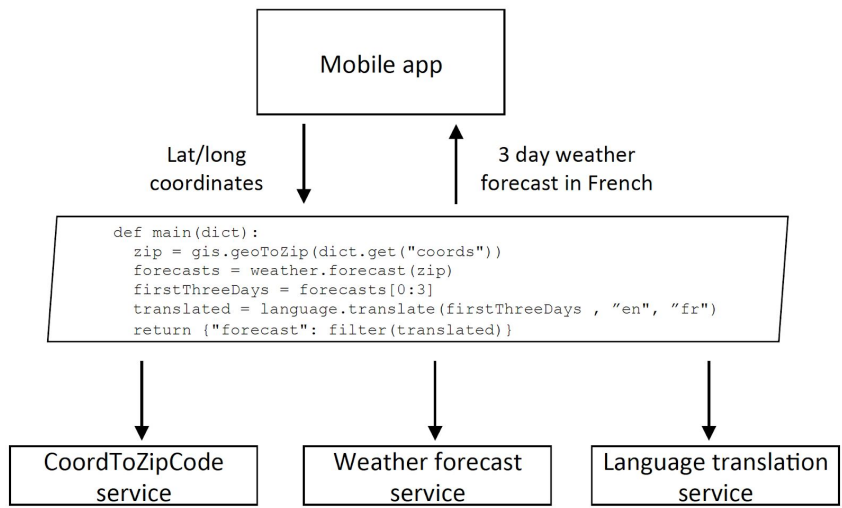

Figure 3: A serverless anti-pattern of offloading API calls from mobile app to backend

Note that the main function in Figure 3 is acting as an orchestrator that is waiting for a response from a function before invoking another, thus incurring a cost of execution while the function is basically waiting for I/O. Such a pattern of programming is referred to as a serverless anti-pattern.

The serverless programming approach would be (c.f. Figure 4) is to encapsulate each API call as serverless function, and the chain the invocation of these functions in a sequence. The sequence itself behaves as a composite function.

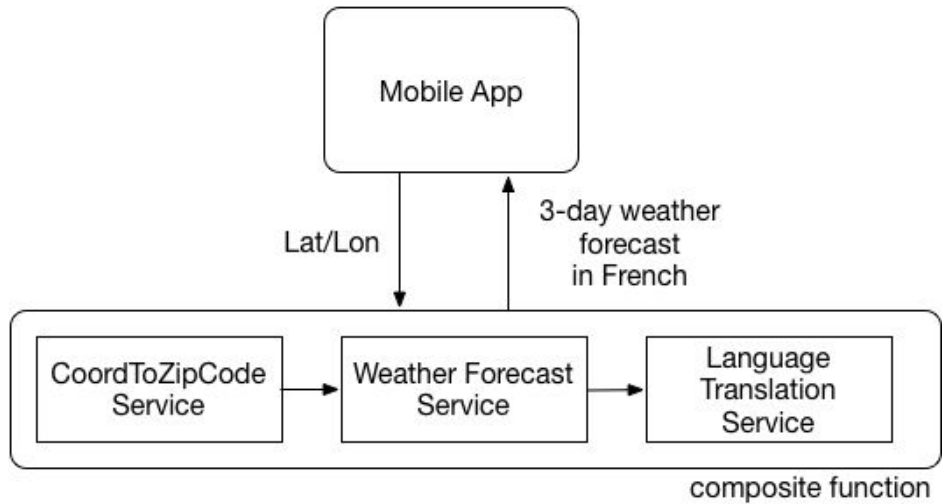

Figure 4: Offloading API calls and glue logic from mobile app to backend

More complex orchestrations can use technologies like AWS Step Functions and IBM Composer to prevent serverless anti-patterns but may incur additional costs due to the services.

## Use Case 3: Map-Reduce style analytics

PyWren [7] (c.f. Figure 5) is a Python based system that utilizes the serverless framework to help users avoid the significant development and management overhead of running MapReduce jobs. It is able to get up to 40 TFLOPS peak performance from AWS Lambda, using AWS S3 for storage and caching. A similar reference architecture has been proposed by AWS Labs[23].

PyWren exemplifies a class of use cases that uses a serverless platform for highly parallel analytics workloads.

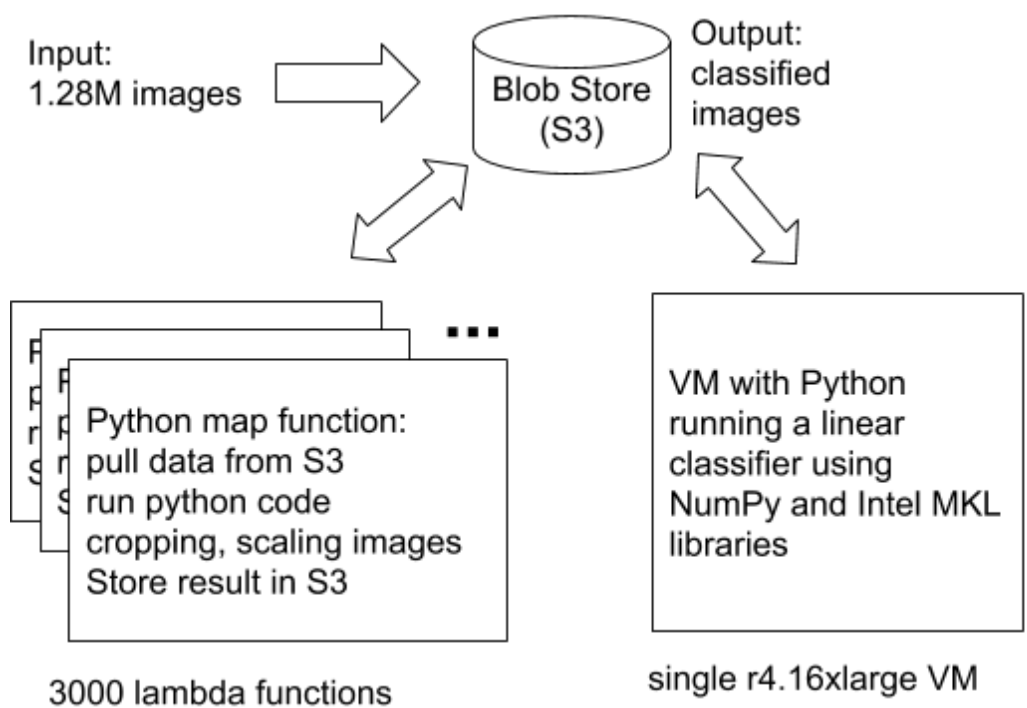

Figure 5: Map + monolithic Reduce PyWren example implementing ImageNet Large Scale Visual Recognition Challenge

---

[23] https://github.com/awslabs/lambda-refarch-mapreduce

# Use Case 4: Multi-tenant cloud services

A Cloud Guru is a company whose mission is to provide users with cloud training that includes videos. An important part of their business model is providing service on-demand and optimizing delivery cost. Their usage patterns are unpredictable and may change depending on holidays or if they do promotions. They need to be able to scale and to isolate users for security reasons while providing for each user backend functionality such as payment processing or sending emails.

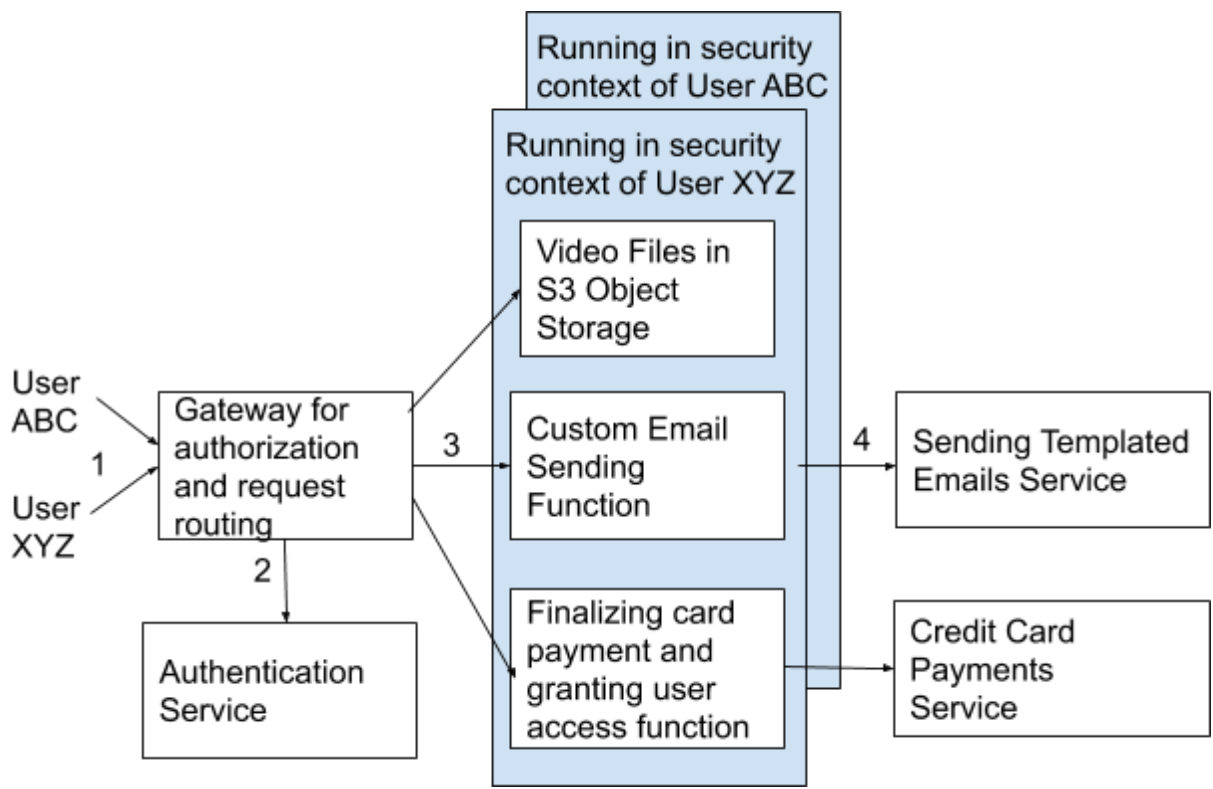

Figure 6: Requests are authenticated and routed to a custom function that runs in isolation and with the user's context.

They achieve this by leveraging cloud services and serverless computing to build a multi-tenant, secure, highly available, and scalable solution that can run each user specific code as serverless functions[24]. This dramatically simplifies how a multi-tenant solution is architected as shown in Figure 6 below. A typical flow starts with a user

---

[24] https://read.acloud.guru/serverless-the-future-of-software-architecture-d4473ffed864

> making a request (1) from a frontend application (web browser). The request is authenticated (2) by using an external service and then sent either to a cloud service (such as object store to provide video files) or (3) to a serverless function. The function makes necessary customizations and typically invokes other functions or (4) cloud services.

## Challenges and Limitations

Serverless computing is a large step forward, and is receiving a lot of attention from industry and is starting to gain traction among academics. Changes are happening rapidly and we expect to see different evolutions of what is serverless and FaaS. While there are many immediate innovation needs for serverless [4, 19, 24], there are significant challenges that need to be addressed to realize full potential to serverless computing. Based on discussions during a series serverless workshops organized by the authors[25], and several academic [9] and industrial surveys[26], we outline the following challenges:

**Programming models and tooling:** since serverless functions are running for shorter amounts of time there will be multiple orders of magnitude more of them that compose applications (e.g. SparqTV[27], a video streaming service runs more than 150 serverless functions).   This however, will make it harder to debug and identify bottlenecks. Traditional tools that assumed access to servers (e.g. root privilege) to be able to monitor and debug applications are not applicable in for serverless applications, and new approaches are needed. Although some of these tools are starting to be available, Higher level development IDEs, tools for orchestrating and composing applications will be critical. In addition, the platform may need to be extended with different recovery semantics, such as at-least-once or at-most-once, or more sophisticated concurrency semantics, such as atomicity where function executions are serialized. As well, refactoring functions (e.g., splitting and merging them), and reverting to older versions, need to be fully supported by the serverless platform.  While these problems have received a lot of attention from the industry and academia [16], there is still a lot of progress to be made.

---

[25] https://www.serverlesscomputing.org/workshops/
[26] https://www.digitalocean.com/currents/june-2018/
[27] https://www.serverlesscomputing.org/wosc3/#sparqtv

**Lack of standards and vendor lock-in:** Serverless computing and FaaS are new and quickly changing and currently there is no standards. As the area matures standards can be expected to emerge. In the meantime, developers can use tools and frameworks that allow the use of different serverless computing providers interchangeably.

## Research Opportunities

Since serverless is a new area, there are many opportunities for the research community to address. We highlight the following:

**System level research opportunities:** A key differentiator of serverless is the ability to scale to zero, and not charge the customers for idle time. Scaling to zero, however, leads to problems of cold starts, particularly for functions with customized library requirements [17]. Techniques to minimize the cold start problem while still scaling to zero are critical. A more fundamental question which is currently being asked is if containers are the right abstractions for running serverless applications and whether abstractions with smaller footprints such as unikernels are more suitable.

**Legacy code in serverless:** Serverless application designs are fundamentally different from typical legacy applications. The economical value of existing code represents a huge investment of countless hours of developers coding and debugging software. One of the most important problems may be to what degree existing legacy code can be automatically or semi-automatically decomposed into smaller-granularity pieces to take advantage of these new economics.

**Stateful serverless:** Current serverless platforms are mostly stateless, and it is an open question if there will be inherently stateful serverless applications in the future with different degrees of quality-of-service without sacrificing the scalability and fault tolerance properties.

**Service level agreements (SLA):** Serverless computing is poised to make developing services easier, but providing QoS guarantees remains difficult [5, 17]. While the serverless platform needs to offer some guarantees of scalability, performance, and availability, this is of little use if the application relies on an ecosystem of services, such as identity providers, messaging queues, and data persistence, which are outside the control of the serverless platform. To provide certain QoS guarantees, the serverless platform needs to communicate the required QoS requirements to the dependent components. Furthermore, enforcement may be needed across functions and APIs, through the careful measurement of such services, either through a third party evaluation system, or self-reporting, to identify the bottlenecks.

**Serverless at the edge:** There is a natural connection between serverless functions and edge computing as events are typically generated at the edge with the increased adoption of IoT and other mobile devices. iRobot's use of AWS Lambda and Step Functions for image recognition was described by Barga as an example of an inherently distributed serverless application [18]. Recently, Amazon extended its serverless capabilities to an edge based cloud environment by releasing AWS Greengrass. Consequently, the code running at the edge, and in the Cloud may not just be embedded but virtualized to allow movement between devices and cloud. That may lead to specific requirements that redefine cost. For example, energy usage may be more important than speed.

**New serverless applications:** The serverless programming model is inherently different, but that should be a motivation to think about building -- or rebuilding -- new and innovative solutions that tap into its what it can provide. Pywren [7], ExCamera [8], high performance computing, numerical analysis, and AI chatbots are but some examples of how scientists are using serverless to come up with new solutions and applications.

# Conclusion

Serverless computing is an evolution in cloud application development, exemplified by the Function-as-a-Service model where users write small functions which are then managed by the cloud platform. This model has proven useful in a number of application scenarios ranging from event handlers with bursty invocation patterns, to compute-intensive big data analytics. Serverless computing lowers the bar for developers by delegating to the platform provider much of the operational complexity of monitoring and scaling large scale applications. However, the developer now needs to work around limitations on the stateless nature of their functions, and understand how to map their application's SLAs to those of the serverless platform and other dependent services. While many challenges remain, there have been rapid advances in the tools and programming models offered by industry, academia, and open source projects.